\documentclass[
    preprint,12pt
]{elsarticle}
\usepackage{graphicx} % Required for inserting images
\usepackage[top=3cm, bottom=3cm, left=3cm, right=3cm]{geometry}
\usepackage{amsmath}
\usepackage{amsfonts}
\usepackage{amssymb}
\usepackage{graphicx}
\graphicspath{{FiguresPDF/}}
\usepackage[colorlinks,citecolor=blue,linktoc=all,linkcolor=cyan]{hyperref}
\usepackage{svg}
\usepackage{mathtools}
\usepackage{upgreek}
\usepackage{physics}
\usepackage{multirow}
\usepackage{placeins} % To control floating objects (figures and tables)
\usepackage{pgfplots}
\pgfplotsset{compat=1.18}
\usepackage{subcaption}
\usepackage{orcidlink}  
\usepackage{caption}
\usepackage{array}

\usepackage[T1]{fontenc}
\usepackage{dsfont}               % use mathds instead of mathbb for outline fonts
\usepackage{mathrsfs}             % provides mathscr without overwriting mathcal
\usepackage{slashed}              % For Dirac slash notation.
\usepackage{amsbsy}
\usepackage{braket}
\usepackage{tikz}
\usepackage{bbold}

\journal{Physics Letters B}

\begin{document}

\begin{frontmatter}

\title{The Sill distribution for all partial waves}
\author[inst1]{Krzysztof Kyzio\l{} \orcidlink{0000-0002-9290-3607}}
%\email{krzysztof.kyziol@ujk.edu.pl}

\author[inst1,inst2]{Francesco Giacosa \orcidlink{0000-0002-7290-9366}}
%\email{francesco.giacosa@ujk.edu.pl}

\affiliation[inst1]{{Institute of Physics, Jan Kochanowski University, Kielce, Poland}}
\affiliation[inst2]{{Institute of Theoretical Physics, Goethe University, Frankfurt am Main, Germany}}

\date{\today}

\begin{abstract}
    In high-energy physics, short-lived resonances appearing in scattering processes are described by energy-dependent spectral functions. The relativistic Breit--Wigner distribution (rBW), obtained as a relativistic extension of the standard Breit--Wigner form, is commonly employed owing to its simplicity. However, when the lower-energy threshold lies sufficiently close to the resonance peak, sizable deviations from the rBW line shape are expected. The Sill distribution provides a simple way to incorporate the physical threshold while preserving exact normalization, but in its original formulation it is inherently restricted to $s$-wave decays. Here, we extend the Sill distribution to arbitrary partial waves. The correct $\ell$-wave threshold behaviour is implemented while exact normalization is retained through a simple modification of the mass term of the propagator. As a benchmark, the $p$-wave Sill provides a very good description of the $\Delta(1232)^{++}$ line shape up to 1.35 GeV and simultaneously reproduces its on-shell mass and width and its complex pole position, in agreement with independent determinations compiled by the PDG.
\end{abstract}


\end{frontmatter}

\section{Introduction}
    In quantum mechanics (QM), the formation of unstable and short-lived states called resonances is very common. As such, a resonance is not an eigenstate of the theory's Hamiltonian and does not have a definite energy. A very useful phenomenological description of the energy probability distribution is given by the famous Breit-Wigner formula   \cite{Breit:1936zzb,Jackson:1964zd}: 
\begin{equation}
    \label{eq:BW_distribution}
      \rho_{\mathrm{BW}}(E) = \frac{1}{\pi}\frac{  \frac{\Gamma_{\mathrm{BW}}}{2} }{
        (E-M_{\mathrm{BW}})^2+\frac{ \Gamma_{\mathrm{BW}}^2}{4}} \; , 
\end{equation}
where $M_{\mathrm{BW}}$ corresponds to the peak of the energy and $\Gamma_{\mathrm{BW}}=1/\tau$ to the width, that is, the inverse of its lifetime. 
This expression is correctly normalized: $\int_{-\infty}^{+\infty}dE\rho_{BW}(E) =1$. However, $\rho_{\mathrm{BW}}(E)$ has no minimal energy, which cannot be true in a physical system. If the left threshold, denoted as $E_{\mathrm{th}}$, is sufficiently far away from the peak $(\Gamma_{\mathrm{}BW} \ll M_{\mathrm{BW}}-E_{\mathrm{th}})$, threshold effects can be neglected over the resonance
region. The usual quasi-stationary-state approximation then applies,
leading to
the Breit--Wigner energy distribution and, in turn, to an exponential decay law; see e.g. Ref. \cite{Sakurai:2011zz}. 
The fact that the vast majority of unstable quantum states follow an exponential decay (whenever it can be directly measured) is indirect but convincing evidence that the BW approximation is extremely good.  Conversely, the left-energy threshold $E_{\mathrm{th}}$ causes a deviation from the exponential decay law at late times in the form of a power law, as Khalfin \cite{khalfin,ghirardi} recognized long ago and as Refs. \cite{Rothe:2006rma,Giacosa:2025knr} verified experimentally on fluorescence decays. 

In a relativistic context, typical of high-energy physics, short-lived particles with lifetimes of the order of $10^{-22}$ s populate the Particle Data Group (PDG) compilation \cite{ParticleDataGroup:2026lqr}. Experimentally, such states appear as resonant structures in scattering or decay processes and are characterized by a finite mass distribution rather than by a sharply defined energy. These resonances are not asymptotic states of the underlying quantum field theoretical (QFT) description. A commonly employed relativistic extension of the BW distribution (rBW) is expressed in terms of the squared invariant mass $s=E^2$  and reads \cite{Bohm:2004zi,Belle-II:2018jsg,Willenbrock:2022smq}
\begin{equation}
\label{eq:rBW_distribution}
\rho_{\mathrm{rBW}}(s) =
\frac{1}{\pi}
\frac{M_{\mathrm{rBW}}\Gamma_{\mathrm{rBW}}}
{(s-M_{\mathrm{rBW}}^2)^2+
M_{\mathrm{rBW}}^2\Gamma_{\mathrm{rBW}}^2} \text{ ,}
\end{equation}
where $M_{\mathrm{rBW}}$ and $\Gamma_{\mathrm{rBW}}$ denote the rBW mass and width, respectively. (They are often simply referred to as the BW mass and width when the relativistic context is clear.) As in the nonrelativistic case, the rBW distribution has been extremely successful in the description of a wide range of resonances.

Yet, similarly to the BW distribution, Eq.~(\ref{eq:rBW_distribution}) does not contain a physical lower bound. For sufficiently broad resonances, or for states lying close to a decay threshold, threshold effects may become important and can be directly observed in experimental spectral functions. A prominent example is provided by the vector and axial-vector spectral functions reconstructed from hadronic $\tau$ decays by ALEPH, displaying the $\rho$ and the broad $a_1$ structures \cite{ALEPH:2005qgp}. Other notable examples refer to resonances close to the $K\bar K$ threshold, such as $a_0(980)$, $f_0(980)$, and $\phi(1020)$, where the measured $K\bar{K}$ line shape cannot in general be represented by a simple BW form \cite{Flatte:1976xv,Baru:2004xg,BESIII:2020ctr}. Moreover, the normalization of Eq.~(\ref{eq:rBW_distribution}) formally refers to $-\infty<s<\infty$, whereas the physical variable satisfies $s \geq s_{\mathrm{th}}= E_{\mathrm{th}}^2\geq 0$.

More advanced rBW-type parametrizations therefore replace the constant width by an energy-dependent quantity, $\Gamma_{\mathrm{rBW}}\rightarrow\Gamma(s)$ which can incorporate the appropriate threshold and phase-space behaviour \cite{Asner:2025resonances,ALICE:2018qdv}. In this case, however, the simple automatic normalization of Eq.~(\ref{eq:rBW_distribution}) is generally lost, and an overall normalization has to be fixed separately. Moreover, as in Ref. \cite{ALICE:2018qdv}, this modification is retained only in the denominator, while the numerator is kept constant, in order to guarantee that the energy distribution is normalizable. A systematic derivation within QFT, in which the energy dependence follows from the imaginary part of the self-energy, provides the proper analytic and threshold structure \cite{Peskin:1995ev}. The resulting spectral functions, however, may involve rather complicated expressions and can be inconvenient for direct implementation in fits to experimental data.

An alternative to the rBW distribution that incorporates threshold effects, remains exactly normalized, and retains a simple analytic form is the Sill distribution proposed in Ref.~\cite{Giacosa:2021mbz} (see also Ref.~\cite{Giacosa:2023xti}):
\begin{equation}
    \label{eq:Sill_distribution}
    \rho_{\mathrm{Sill}}(s)
    =
    \frac{1}{\pi}
    \frac{\widetilde{\Gamma}\sqrt{s-s_{\mathrm{th}}}}
    {(s-M_{\mathrm{Sill}}^2)^2+
    \widetilde{\Gamma}^{\,2}(s-s_{\mathrm{th}})}
    \,\theta(s-s_{\mathrm{th}})\; .
\end{equation}
The Sill distribution explicitly contains the lower energy threshold $s_{\mathrm{th}}$. For a two-body decay of a resonance $R$ into particles $A$ and $B$ ($R \rightarrow AB$), one has
$s_{\mathrm{th}}=E_{\mathrm{}th}^2=(m_A+m_B)^2$.
The remaining parameters are the mass $M_{\mathrm{Sill}}$ and the rescaled width $\widetilde{\Gamma}$. The latter is the asymptotic value of the energy-dependent decay function $\Gamma(s)$, in turn leading to the on-shell Sill width given by:

\begin{equation}  \Gamma(s)=\widetilde{\Gamma} \frac{\sqrt{s-s_{\mathrm{th}}}}{\sqrt{s}} \rightarrow \Gamma_{\mathrm{Sill}}
    =\Gamma(M_{\mathrm{Sill}}^2)=
    \widetilde{\Gamma}
    \frac{\sqrt{M_{\mathrm{Sill}}^2-s_{\mathrm{th}}}}
    {M_{\mathrm{Sill}}}\; .
    \label{eq:sqveGamma}
\end{equation}
At threshold, the spectral function vanishes continuously and displays the characteristic ``sill-like'' square-root rise. A particularly useful feature is the exact normalization:
\begin{equation}
    \int_{s_{\mathrm{th}}}^{\infty} ds\,
    \rho_{\mathrm{Sill}}(s)=1
\end{equation}
for arbitrary values of $s_{\mathrm{th}}$, $M_{\mathrm{Sill}}$, and $\widetilde{\Gamma}$ (with $M_{\mathrm{Sill}}^2>s_{\mathrm{th}}$). 

Applications of the Sill to various resonances were presented in
Refs.~\cite{Giacosa:2021mbz,Giacosa:2023xti}. Owing to its simplicity and versatility, it has subsequently been employed in theoretical studies, e.g.~Refs. \cite{Winney:2022tky,Sarti:2023wlg,Yildirim:2023znd}, as well as in experimental analyses by the ALICE and NA61/SHINE Collaborations \cite{ALICE:2025aur,ALICE:2025byl,NA61SHINE:2023azp}.
Just as (r)BW, the Sill distribution can be in principle applied to any decay. Yet, while the presence of the threshold is taken into account, the Sill function is inherently s-wave because of the close-to-threshold dependence $\sqrt{s-s_{\mathrm{th}}} \propto k_{CM}$, with $k_{CM} = |\vec{k}_A|= |\vec{k}_B|$ being the modulus of the three-momentum of the outgoing particles in the rest frame of $R$.

It is then natural to extend the Sill to any $\ell$-th wave decay, while retaining its simplicity and correct normalization: this is the goal of the present work. We first establish the general shape of $\rho_{\mathrm{Sill}}^{(\ell)}(s)$ for any $\ell$-wave, whose straightforward algebraic expression can be easily implemented in numerical simulations. Finally, we concentrate on the p-wave Sill and, as a specific notable example, on the decay and properties of the baryonic resonance $\Delta^{++} \rightarrow p \pi^+$, which is strongly affected by the nearby decay threshold. A successful fit of the p-wave Sill to SAID proton-pion scattering data makes these effects explicit.

The paper is organized as follows: in Sec. 2 we derive the Sill distribution for any partial wave, with special attention to the p-wave. In Sec. 3 the p-wave Sill is applied to the resonance $\Delta^{++}(1232)$. In Sec.  4 we briefly present our conclusions.

\section{Derivation of the $\ell$-th wave Sill distribution}

The central quantity for a resonance is (the scalar part of) the propagator of an unstable state (e.g  Ref. \cite{Peskin:1995ev,Matthews:1958sc}): 
\begin{equation}
        \label{eq:propagator_definition}
        G(s) = \frac{1}{s-M_0^2+\Pi (s) + i\varepsilon} \; ,
    \end{equation}
where $M_0$ is the bare mass and $\Pi(s)$ is the self-energy function determined by the specific coupling of the resonant state with its decay products; see
Ref.~\cite{Giacosa:2007bn} for a discussion in terms of spectral
functions and Refs.~\cite{Mai:2022eur,Asner:2025resonances} for recent reviews.

Within QFT models, the self-energy $\Pi(s)$ is fixed by the interaction Lagrangian and the adopted regularization and renormalization prescription. For scalar resonances, the easiest interaction is of the type $\mathcal{L}=gRAB$, see, e.g., Refs.~\cite{Oller:1997ti,Boglione:1997aw,Giacosa:2007bn}; related coupled-channel descriptions of near-threshold scalar resonances can be found, e.g., in Ref.~\cite{Baru:2003qq}. For derivative interactions, such as $\mathcal{L}=gR \, \partial_{\mu}A \, \partial^{\mu}B$, see, e.g., Refs.~\cite{Black:2006mn,Giacosa:2008st}. Examples involving vector or tensor fields for the decaying state $R$ and/or for its decay products $A$ and $B$ can be found in Refs.~\cite{Molina:2008jw, Molina:2009ct,Coito:2019cts}. Analogous constructions for baryonic resonances are discussed in e.g. Refs.~\cite{Doring:2009uc,Gegelia:2016xcw}.

Alternatively, resonance dynamics, together with modelling of the loop function $\Pi(s)$, can be formulated in terms of the Lee-Friedrichs interaction Hamiltonian \cite{Lee:1954iq,friedrichs1948}, in which a discrete bare state is coupled to a continuum of states; see also Refs.~\cite{Civitarese:2004xbt,Ordonez:2001zz,Xiao:2016wbs,Xiao:2016dsx,Giacosa:2020tha} and references therein. Relativistic extensions of this approach have been also put forward \cite{Fuda:1990nc,Giacosa:2010br,Zhou:2020vnz,Giacosa:2021hgl,Zhou:2020moj}.

From the propagator in Eq. (\ref{eq:propagator_definition}), important features of the system emerge. The most significant relation is the link between the spectral function and the imaginary part of the propagator:
\begin{equation}
\label{eq:spectral_function_propagator_relation}
        \rho (s) = -\frac{1}{\pi} \Im \left[ G(s)  \right] \; .
\end{equation}
For our purposes, also the scattering amplitude $\mathcal{M} (s)$, defined as:
    \begin{equation}
        \label{eq:scattering_amplitude_Lee_Freidrichs}
        \mathcal{M}(s) = \frac{\Im \left[ \Pi (s)  \right]}{s-M_0^2+\Pi (s) + i\varepsilon} \; ,
    \end{equation}
plays an important role in the modelling of the self-energy function.
In general, the unitarity condition for the scattering amplitude (see e.g. Ref. \cite{Mandelstam:1958xc,Pelaez:2015qba,Oller:2019opk}), which corresponds to the S-matrix for the elastic case, leads to the following expression:
\begin{equation}
        \label{eq:scattering amplitude}
        \mathcal{M}(s) = e^{i\delta(s)} \sin \left[ \delta (s) \right] \; .
\end{equation}
Above, $\delta(s)$ denotes the phase shift as a function of the squared
invariant mass. Near threshold, the Wigner threshold law implies that,
for a partial wave with orbital angular momentum $\ell$, the phase shift
behaves as \cite{Wigner:1948zz,Peters:2004qw}
\begin{equation}
\label{eq:phase_angle_low_energy}
    \delta(s) \propto k_{\mathrm{CM}}^{\,2\ell+1}\; ,
\end{equation}
where $k_{\mathrm{CM}}$ is the centre-of-mass momentum and $\ell$ denotes
the orbital angular momentum quantum number of the partial wave.
 
 We recall that, in a two-body decay $R \rightarrow AB$, the momentum $k_{\mathrm{CM}}$ takes the explicit expression \cite{ParticleDataGroup:2026lqr}: 
\begin{equation}
k_{\mathrm{CM}} = \sqrt{\frac{(s-(m_A+m_B)^2)(s-(m_A-m_B)^2)}{4s}}  \simeq \tfrac{1}{2}(s-s_{\mathrm{th}})^{1/2} \text { .}
     \label{approx}
 \end{equation}
 The term on the r.h.s. is exact for $m_A = m_B$ but represents a very good approximation even when $m_A \neq m_B$ for all practical cases considered. The Sill distribution makes use of this approximation, which considerably simplifies many analytic expressions.

 Combining Eqs. (\Ref{eq:scattering_amplitude_Lee_Freidrichs}), (\Ref{eq:scattering amplitude}), (\Ref{eq:phase_angle_low_energy}), and (\ref{approx}), the requirement for the self-energy function for an arbitrary $\ell$ reads:
    \begin{equation}
        \label{eq:self_energy_function_constraint}
        \Im \left[ \Pi_{\ell}(s) \right] \propto ( s- s_{\mathrm{th}})^{\ell +1/2} \; .
    \end{equation}
In this way, we take into account the left-energy threshold. 
On the other hand, at high energies, the self-energy function is assumed to regain the Breit-Wigner shape, thus: 
\begin{equation}
        \label{eq:self-energy_function_high_energies}
        \Im \left[ \Pi_{\ell} (s) \right] = \sqrt{s} \; \widetilde{\Gamma}  \text{  for large } s \text { .}
\end{equation}
Eq.~(\ref{eq:self_energy_function_constraint}) and
Eq.~(\ref{eq:self-energy_function_high_energies}) can be fulfilled simultaneously by choosing
\begin{equation}
    \label{eq:self-energy_function_model}
    \Im \left[\Pi_{\ell}(s)\right]
    =
    \frac{\widetilde{\Gamma}
    \left(s-s_{\mathrm{th}}\right)^{\ell+1/2}}
    {\left(s-s_0\right)^{\ell}} \, \theta(s-s_{\mathrm{th}}) \; ,
\end{equation}
where $\widetilde{\Gamma}$ controls the asymptotic value of the width, while
$s_0<s_{\mathrm{th}}$ introduces a subthreshold scale governing the
near-threshold suppression and the transition toward the asymptotic regime\footnote{The limiting
case $s_{\mathrm{th}}=0$ is instructive: for $\ell>0$ one must
retain $s_0<0$, since setting $s_0=0$ would cancel the $\ell$-dependent
suppression and reduce Eq.~(\ref{eq:self-energy_function_model}) to
$\Im[\Pi_{\ell}(s)]\propto\sqrt{s}$, independently of $\ell$.}.

\begin{figure}[!htbp]
    \centering
    \includegraphics[width=0.6\linewidth]{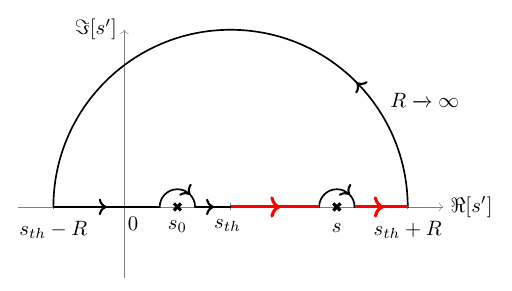}
    \caption{Contour of the integration. The part marked with red colour corresponds to the interval of integration from Eq. (\ref{eq:real_part_self-energy}). The final result is obtained in the limit as $R \xrightarrow{} \infty$.}
    \label{fig:integration_contour}
\end{figure}

Once the imaginary part of $\Pi_{\ell} (s)$ is modelled, the real part can be obtained via dispersion relations (also known as the Kramers-Kr\"{o}nig relations \cite{deL.Kronig:26}):
    \begin{equation}
        \label{eq:real_part_self-energy}
            \Re  \left[ \Pi_{\ell} (s) \right] = \frac{1}{\pi} P\int_{s_{\mathrm{th}}}^{\infty} \mathrm{d}s' \frac{\Im \left[ \Pi_{\ell}(s') \right]}{s'-s} = \frac{1}{\pi} P\int_{s_{\mathrm{th}}}^{\infty} \mathrm{d}s' \frac{\widetilde{\Gamma} \left( s'-s_{\mathrm{th}}\right) ^{\ell+1/2}}{\left(s'-s_0 \right)^{\ell}} \frac{1}{s'-s}  \; ,
    \end{equation}
    that are fulfilled for any analytic complex function.  \smallskip \newline
    The integral is evaluated using standard complex analysis methods with the contour of the integration presented in Fig. \ref{fig:integration_contour}. The fact that the integrand takes purely imaginary values below the threshold allows us to write:
    \begin{equation}
        \label{eq:real_part_self-energy_trick}
        \Re \left[ \Pi_{\ell} (s) \right]  =  \frac{1}{\pi} \Re  \left[ P \int_{-\infty}^{\infty} \mathrm{d}s' \frac{\widetilde{\Gamma} \left( s' - s_{\mathrm{th}}\right) ^{\ell + 1/2}}{\left(s'-s_0 \right)^{\ell}  (s'-s)} \right] .
\end{equation}
The contribution of the large semicircle is formally a divergent (or regularization-dependent) constant that can be reabsorbed into the bare mass parameter $M_0$. The `physical' Sill mass is defined by:
    \begin{equation}
        \label{eq:Sill_physical_mass}
        M_{\mathrm{Sill}}^2-M_0^2+\Re \left[ \Pi_{\ell} (M_{\mathrm{Sill}}^2) \right] = 0 \text{ .}
    \end{equation}
Hence, for a given $\ell$, the integral of Eq. (\Ref{eq:real_part_self-energy_trick}) takes the expression below:
\begin{equation}
        \label{eq:real_part_self-energy_final_result}
        \Re \left[ \Pi_{\ell} (s) \right] = \widetilde{\Gamma} \sum_{n=0}^{\ell -1} \binom{\ell + \frac{1}{2}}{n} \left( -1 \right) ^{\ell - n} \, \frac{\left( s_{\mathrm{th}}-s_0 \right) ^{\ell - n +1/2}}{\left( s-s_0 \right) ^{\ell - n}} \; .
    \end{equation}
For the particular p-wave case ($\ell =1$), Eq. (\ref{eq:real_part_self-energy_final_result}) reduces to a relatively simple formula:
\begin{equation}
\label{eq:p_wave_real_part_self-energy}
        \Re \left[ \Pi _{\ell}(s) \right]  = - \frac{\widetilde{\Gamma} \left( s_{\mathrm{th}} -s_0 \right) ^{3/2}}{ s-s_0} .
    \end{equation}
Hence, combining Eqs. (\ref{eq:propagator_definition}), (\ref{eq:spectral_function_propagator_relation}), (\ref{eq:self-energy_function_model}), (\ref{eq:Sill_physical_mass}), and (\ref{eq:p_wave_real_part_self-energy}) we obtain the complete expression for the p-wave Sill distribution:
\begin{equation}
\label{eq:p_wave_Sill_distribution}
        \rho_{\mathrm{Sill}}^{(p)}(s) = \frac{1}{\pi}\frac{\frac{\widetilde{\Gamma}(s-s_{\mathrm{th}})^{3/2}}{s-s_0}}{\left(s-M_{\mathrm{Sill}}^2-\frac{\widetilde{\Gamma}\left( s_{\mathrm{th}}-s_0 \right) ^{3/2}}{s-s_0}+\frac{\widetilde{\Gamma} \left( s_{\mathrm{th}}-s_0 \right) ^{3/2}}{M_{\mathrm{Sill}}^2-s_0} \right)^2 + \left( \frac{\widetilde{\Gamma}(s-s_{\mathrm{th}})^{3/2}}{s-s_0} \right) ^2} \; .
    \end{equation}
This formula can be easily implemented numerically. 
    
Finally, for any $\ell$, the Sill distribution takes the following form:
\begin{equation}
    \label{eq:l_wave_Sill_distribution}
    \rho_{\mathrm{Sill}}^{(\ell)}(s)  =\frac{1}{\pi}\frac{\frac{\widetilde{\Gamma}(s-s_{\mathrm{th}})^{\ell +1/2}}{(s-s_0)^{\ell}}}{\left( s-M_{\mathrm{Sill}}^2+\Re \left[ \Pi _{\mathrm{\ell}}(s) \right] - \Re \left[ \Pi _{\mathrm{\ell}}(M_{\mathrm{Sill}}^2) \right] \right)^2 + \left( \frac{\widetilde{\Gamma}(s-s_{\mathrm{th}})^{\ell +1/2}}{(s-s_0)^{\ell}} \right) ^2} \; ,
\end{equation}
with $\Re \left[ \Pi _{\mathrm{\ell}}(s) \right]$ defined by Eq. (\ref{eq:real_part_self-energy_final_result}).
Even if the complexity increases with $\ell$, it remains an algebraic expression that can be straightforwardly included in numerical calculations and fits to experimental data. The Sill decay width that generalizes Eq. (\ref{eq:sqveGamma}) reads:
\begin{equation}  \Gamma(s)=    \frac{\widetilde{\Gamma}
    \left(s-s_{\mathrm{th}}\right)^{\ell+1/2}}
    {\sqrt{s} \left(s-s_0\right)^{\ell}}   \rightarrow \Gamma_{\mathrm{Sill}}
    =\Gamma(M_{\mathrm{Sill}}^2)=
    \frac{\widetilde{\Gamma}
    \left(M_{\mathrm{Sill}}^2-s_{\mathrm{th}}\right)^{\ell+1/2}}
    {M_{\mathrm{Sill}} \left(M_{\mathrm{Sill}}^2-s_0\right)^{\ell}}\; .
    \label{eq:lwaveGamma}
\end{equation}
\section{Application to the $\Delta (1232)^{++}$}

The resonance $\Delta (1232)^{++}$ is an optimal test case for the p-wave Sill. Namely, the threshold reads  $\sqrt{s_{\mathrm{th}}}=M_p+M_{\pi^+}=1077.842$ MeV (we neglect the error), while
a weighted average of the independent rBW determinations reported in the PDG gives
$M_{\mathrm{rBW}}=1230.8\pm0.2$ MeV and
$\Gamma_{\mathrm{rBW}}=111.5\pm0.4$ MeV, thus
\begin{equation}
    \frac{\Gamma_{\mathrm{rBW}}}
    {M_{\mathrm{rBW}}-\sqrt{s_{\mathrm{th}}}}
    \simeq 0.73 \ ,
    \nonumber
\end{equation}
which is close to $1$: sizable threshold effects are expected. 

The resonance $\Delta (1232)^{++}$ dominates in the low energy regime in the P33 channel in $p$-$\pi^{+}$ scattering. The partial wave amplitudes corresponding to that channel have been extracted from total cross section measurements (see e.g. Refs. \cite{Troka:1966zz, Carter:1968zza, Carter:1971tj, Pedroni:1978it}) by Workman et al. \cite{Workman2012} and can be found in the SAID Database \cite{SAID}. The relation between the cross section $\sigma$ and the phase angle $\delta$ from Eq. (\Ref{eq:scattering amplitude}) is expressed as follows \cite{haskins1985}:
\begin{equation}
    \sigma_{\mathrm{P33}} (s) = \frac{8\pi}{k^2} \, \sin ^2 \left[ \delta_{\mathrm{P33}} (s) \right] \; .
\end{equation}
Moreover, the scattering amplitude is modelled using Eq. (\ref{eq:scattering_amplitude_Lee_Freidrichs}):
\begin{equation}
    M_{\mathrm{P33}}(s) =\frac{\frac{\widetilde{\Gamma}(s-s_{\mathrm{th}})^{3/2}}{s-s_0}}{ s-M_{\mathrm{Sill}}^2-\frac{\widetilde{\Gamma}\left( s_{\mathrm{th}}-s_0 \right) ^{3/2}}{s-s_0}+\frac{\widetilde{\Gamma} \left( s_{\mathrm{th}}-s_0 \right) ^{3/2}}{M_{\mathrm{Sill}}^2-s_0}  +  \frac{i\widetilde{\Gamma}(s-s_{\mathrm{th}})^{3/2}}{s-s_0}} \; .
\end{equation}
This approach is fully compatible with Eq. (\ref{eq:scattering amplitude}). \smallskip \newline

\begin{figure}[!htb]
    \centering
    \begin{subfigure}{0.44\textwidth}
        \centering
        \includegraphics[width=1\linewidth]{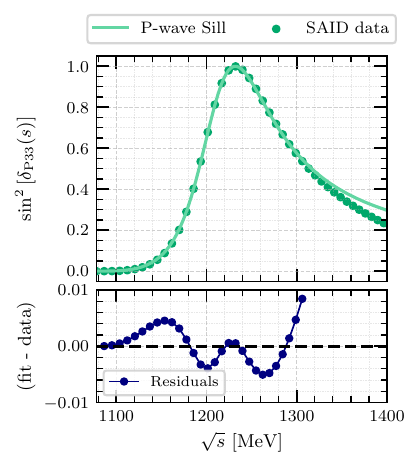}
        \caption{Sine squared of the phase angle.}
        \label{fig:Delta_sine_squared}
    \end{subfigure}
    \hfill
    \begin{subfigure}{0.52\textwidth}
        \centering
        \includegraphics[width=1\linewidth]{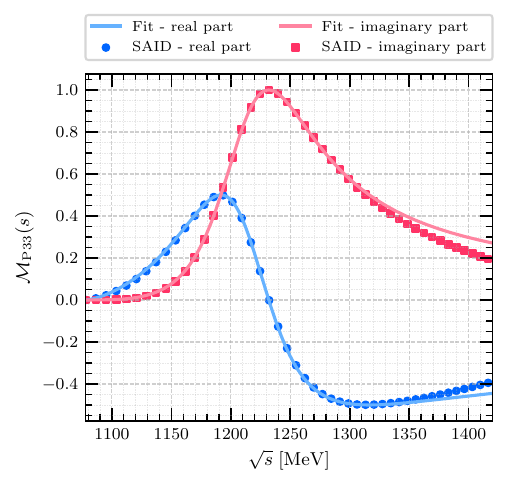}
        \caption{Components of the scattering amplitude.}
        \label{fig:Delta_scattering_amplitude}
    \end{subfigure}
    \caption{Application of the p-wave Sill distribution to $\Delta (1232)^{++}$. Data comes from SAID database. No background term, but $s_0\neq0$.}
    \label{fig:Delta_P-wave_Sill}
\end{figure}

In our treatment, we consider the s-channel $p\pi^+ \rightarrow\Delta^{++} \rightarrow p\pi^+$, in which the resonance $\Delta^{++}$ dominates\footnote{The $\pi^+$ is, strictly speaking, not stable and formally not an asymptotic state; yet, as shown in Ref. \cite{Giacosa:2025fvq} the scattering off unstable states can be justified if the lifetime is much longer than the time-scale of the dominant interaction.}.
Additional nonresonant mechanisms, such as $t$-channel exchanges of the
$f_0(500)$, $f_0(980)$, $\rho^0$, \ldots mesons, or Coulomb repulsion, may
also contribute, in particular in the near-threshold region.
However, in the present description, no explicit background term is
introduced. The parameter $s_0$ provides the $p$-wave Sill with sufficient
flexibility to reproduce the observed near-threshold behaviour and may
effectively absorb smooth contributions that are not resolved separately.
Indeed, the description remains very good close to the threshold, whereas
deviations become visible only at higher energies. At approximately $1.35$ GeV,
contributions from heavier resonances in the $s$ channel become increasingly
relevant. For a general discussion of the inclusion of background terms
consistent with unitarity requirements, see, e.g., Ref.~\cite{Heuser:2024biq}. The overall fit shows that the p-wave Sill describes the data very well up to $\sim1.35$ GeV both for the phase angles and the real and imaginary parts of the amplitude; see Fig. \Ref{fig:Delta_P-wave_Sill}.
The best-fit parameters are presented in Table \Ref{tab:fit-parameters}. We remark that using the s-wave Sill or the rBW  (with or without background) leads to much poorer fits, confirming that the proper behaviour near the threshold is an important element for a correct description of scattering data. 

{
\renewcommand{\arraystretch}{1.5}
\begin{table}[!htb]
    \centering
    \caption{The Sill parameters fitted to scattering amplitudes in the P33 channel.}
    \begin{tabular}{|c|c|}
        \hline
        Parameter & Value \\\hline \hline
        $\sqrt{s_{\mathrm{th}}}$ & 1077.842 MeV (fixed) \\\hline
        $\widetilde{\Gamma}$ & 477.4 $\pm$ 9.3 MeV \\\hline
        $M_{\mathrm{Sill}}$ & 1231.90 $\pm$ 0.14 MeV \\\hline
        $\sqrt{s_0}$ & 941.5 $\pm$ 4.6 MeV \\\hline
    \end{tabular}
    \label{tab:fit-parameters}
\end{table}
}

\begin{figure}[!htb]
    \centering
    \begin{subfigure}{0.47\textwidth}
        \centering
        \includegraphics[width=1\linewidth]{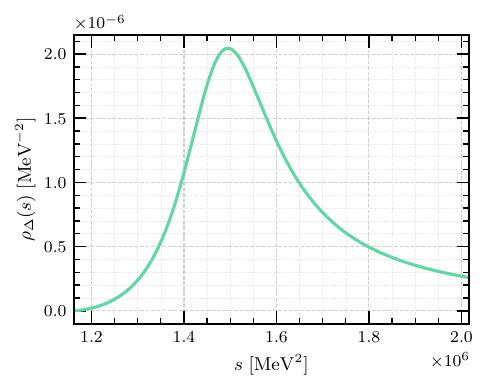}
        %\caption{Plot of the spectral function.}
        \label{fig:Spectral_func}
    \end{subfigure}
    \hfill
    \begin{subfigure}{0.49\textwidth}
        \centering
        \includegraphics[width=1\linewidth]{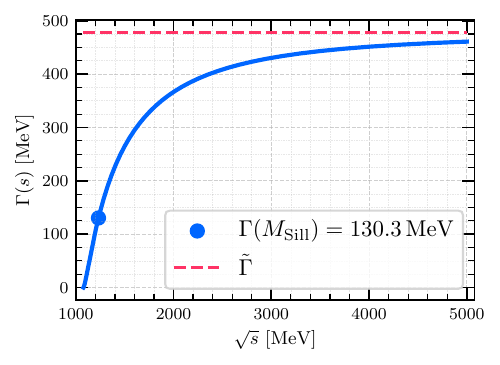}
        %\caption{Plot of the decay width function.}
        \label{fig:Decay_width}
    \end{subfigure}
    \caption{Left panel: p-wave Sill spectral function of the resonance $\Delta (1232)^{++}$. Right panel: energy-dependent Sill decay width of the p-wave Sill distribution ($\ell=1$ in Eq. (\ref{eq:lwaveGamma})). The dot corresponds to $\Gamma_{\mathrm{Sill}}$.}
    \label{fig:P-wave_Sill_basic_plots}
\end{figure}

As an important application of our approach, we extract the position of the pole in the complex plane. This is achieved by setting the denominator of Eq. (\ref{eq:propagator_definition}) equal to 0. 
Table \ref{tab:pole_position} presents the pole positions related to the p-wave Sill compared to those listed in the Particle Data Group (PDG).
{
\renewcommand{\arraystretch}{1.5}
\begin{table}[!htb]
    \centering
    \caption{Position of the pole in the complex plane.}
    \begin{tabular}{|c|c|c|}
        \hline
         & Fit value & PDG value \cite{Bernicha:1995gg} \\\hline \hline
        $\Re \left[ \sqrt{s_{\mathrm{pole}}}\right] $ & 1213.12 $\pm$ 0.22 MeV & 1212.50 $\pm$ 0.24 MeV\\\hline
        -2 $\Im \left[ \sqrt{s_{\mathrm{pole}}}\right] $ & 98.39 $\pm$ 0.42 MeV & 97.37 $\pm$ 0.42 MeV \\\hline
    \end{tabular}
    
    \label{tab:pole_position}
\end{table}
}
\smallskip \newline
Moreover, for the values of the best-fit parameters, both the plot of the invariant mass distribution (see Eq. (\ref{eq:p_wave_Sill_distribution})) and the decay width (see Eq. (\ref{eq:self-energy_function_high_energies})) are presented in Fig. \ref{fig:P-wave_Sill_basic_plots}.
In the right panel of Fig. {\ref{fig:P-wave_Sill_basic_plots}}, one can see that $\widetilde{\Gamma}$ represents the upper limit of the decay width function. However, in the particular case of $\Delta (1232)^{++}$, the region where $\Gamma (s)$ saturates lies far beyond the resonance peak. Hence, in this case, $\widetilde{\Gamma}$ should not be compared directly with Breit-Wigner decay width values from previous analyses, e.g., \cite{Bernicha:1995gg}). The dot corresponding to the mass is, however, in good agreement with the rBW determination. 

\section{Conclusions}

In this work, we have derived the Sill distribution for any $\ell$-th partial wave, describing the proper threshold behaviour at small energies and correctly normalized to unity for any parameter choice. 

The $\ell$-th wave Sill depends on 4 parameters: the mass $M_{\mathrm{Sill}}$, the width parameter $\widetilde{\Gamma}$, the threshold energy $s_{\mathrm{th}}$, and a near-threshold parameter $s_0$, the latter appearing only if $\ell>0$. The s-wave Sill is correctly reproduced for $\ell=0$. In general, the correction to the mass part of the propagator is not a simple constant for $\ell>0$ but is a relatively simple function that can be easily implemented in numerical applications. 

The specific application of the $p$-wave Sill to the $\Delta(1232)^{++}$ resonance shows very good agreement with the experimental data. Both the extracted pole position and the on-shell mass and width are consistent with independent determinations. In future work, the $\ell$-wave Sill can be applied to other hadronic resonances, of particular relevance for ongoing femtoscopy studies \cite{ALICE:2025byl,Fabbietti:2020bfg}. A straightforward extension to multiple decay channels may further allow for more realistic applications in which several thresholds contribute simultaneously.

\section*{Acknowledgments}
\begin{sloppypar}
    The authors thank Vanamali C. Shastry, Christoph Hanhart, and Valentina Mantovani-Sarti
    %from FZ J\"{u}lich (Institute for Advanced Simulation (IAS)) 
    for stimulating discussions. This work was co-financed by the Minister of Science (Poland) under the "Regional Excellence Initiative" program (project no.: RID/SP/0015/2024/01).
\end{sloppypar}

% --- Bibliography ---
\bibliographystyle{elsarticle-num}
\bibliography{bibliography}

\end{document}